\documentclass{llncs}

\usepackage[utf8]{inputenc}
\usepackage{graphicx}
\usepackage{float}
\usepackage{algorithm}
\usepackage{algpseudocode}

\usepackage{apacite}
\long\def\/*#1*/{}

\begin{document}

\title{Domain specific cues improve robustness of deep learning based segmentation of ct volumes}

\author{Marie Kloenne\inst{1,2\star} \and Sebastian Niehaus\inst{1,3\star} \and Leonie Lampe\inst{1} 
 \and Alberto Merola\inst{1} \and Janis Reinelt \inst{1} \and Ingo Roeder\inst{3, 4} \and Nico Scherf\inst{4, 5} } 

\institute{AICURA medical, Bessemerstrasse 22, 12103 Berlin, Germany
\email{firstname.lastname@aicura-medical.com} \and Technische Fakult\"at, Universit\"at Bielefeld, Universit\"atsstrasse 25, 33615 Bielefeld, Germany\and Institute for Medical Informatics and Biometry, Carl Gustav Carus Faculty of Medicine, Technische Universit\"at Dresden, Fetscherstrasse 74, 01307 Dresden, Germany
\and National Center of Tumor Diseases (NCT), Partner Site Dresden, 01307 Dresden, Germany \and Max Planck Institute for Human Cognitive and Brain Sciences, Stephanstrasse 1a, 04103 Leipzig, Germany}
\renewcommand{\thefootnote}{\fnsymbol{footnote}}
\footnotetext[1]{The authors contributed  equally to this paper.}

\maketitle 
%\keywords{Keyword1, Keyword2, Keyword3}

\begin{abstract}
Machine Learning has considerably improved medical image analysis in the past years. Although data-driven approaches are intrinsically adaptive and thus, generic, they often do not perform the same way on data from different imaging modalities. In particular Computed tomography (CT) data poses many challenges to medical image segmentation based on convolutional neural networks (CNNs), mostly due to the broad dynamic range of intensities and the varying number of recorded slices of CT volumes. In this paper, we address these issues with a framework that combines domain-specific data preprocessing and augmentation with state-of-the-art CNN architectures. The focus is not limited to optimise the score, but also to stabilise the prediction performance since this is a mandatory requirement for use in automated and semi-automated workflows in the clinical environment. \\
 The framework is validated with an architecture comparison to show CNN architecture-independent effects of our framework functionality. We compare a modified U-Net and a modified Mixed-Scale Dense Network (MS-D Net) to compare dilated convolutions for parallel multi-scale processing to the U-Net approach based on traditional scaling operations. Finally, we propose an ensemble model combining the strengths of different individual methods. \\
The framework performs well on a range of tasks such as liver and kidney segmentation, without significant differences in prediction performance on strongly differing volume sizes and varying slice thickness.
Thus our framework is an essential step towards performing robust segmentation of unknown real-world samples.
\end{abstract}

\flushbottom

\thispagestyle{empty}

\section*{Introduction}
Spatial characteristics of tumours like size, shape, location or growth pattern are central clinical features. Changes in these characteristics are essential indicators of disease progression and treatment effects. Automated, quantitative assessment of these characteristics and their changes from radiological images would yield an efficient and objective tool for radiologists to monitor the course of the disease. Thus, a reliable and accurate automated segmentation method is desirable to extract spatial tumour and organ characteristics from computed tomography (CT) volumes.\\
In recent years, convolutional neural networks (CNNs) \cite{Krizhevsky12} became the state of the art method for image segmentation, as well as many other tasks in computer vision \cite{Voulodimos18}, such as image classification, object detection and object tracking \cite{Moen2019}. The applications of CNNs are diverse, but the general data handling or preprocessing is often very similar in each case since the feature extraction is performed internally by the CNN itself. Improvements in the application of CNNs for medical image processing often address the neural network architecture, the training algorithm or the use case \cite{Minnemaa2018,Chlebus18}. At the same time, most authors tend to ignore the data handling itself, treating medical images such as CT volumes the same way as grayscale images or RGB images just with additional dimensions.

However, this approach neglects prior information about the specific physical processes that underlie images acquisition and determine image contrast, possibly leading to suboptimal and sometimes inaccurate image analysis. For instance, while most image formats map pixels on relative scales of a few hundred values, voxels in CT volumes are mapped on the Hounsfield scale \cite{Broder2011}, a quantitative mapping of radiodensity calibrated such that the value for air is -1000 Hounsfield Units (HU) and that for water is 0 HU, with values in the human body reaching up to about 2000 HU (cortical bone). Therefore, in contrast to most standard images where pixel intensities themselves might not be meaningful, the actual grey values of CT volumes carry tissue-specific information \cite{Brenner2007}, and special consideration is required to leverage it. 

The tissue-specific information also means, that CT data typically contains a range of values that are not necessarily relevant for a particular diagnostic question \cite{Costelloe13,Harris93}. Thus, when radiologists inspect CT volumes for diagnosis, they typically rely on windowing, i.e. they restrict the range of displayed grey values to focus the image information to relevant values. CNN-based image segmentation frameworks rarely include such potentially essential steps from the expert workflow. They are assuming that the data only has to be normalised and the network will then learn by itself to focus on the relevant image regions.

In this paper, we address the challenges of a clinically meaningful CT volume processing and present a domain specific framework for CNN based image segmentation. The proposed framework is inspired by insights on both the data acquisition process and the diagnostic process performed by the radiologist, addressing in particular the spatial information CT volumes and the use of the HU scale. \\ Our focus is not on the optimisation of the loss function on the whole dataset, but instead on obtaining a robust segmentation quality, independent of the differences in size and shape of the input volumes. For this reason, we also consider the standard deviation of the dice score as a measure of robustness for evaluation. If we use a segmentation model in an automated or semi-automated process in which the result of the segmentation is not directly analysed, particularly strong segmentation errors pose a problem because the user tends to rely on the segmentation model and only analyse the final result of the process. Therefore, our goal is to specifically address the demands of algorithms for CT processing in the clinical environment, where we require algorithms to process each volume consistently and without significant differences in the quality of the output.

 We evaluated the framework with a mixed-scale dense convolutional neural network (MS-D Net) \cite{Pelt2017} with dilated convolutions and the nnU-Net \cite{Isensee2018} with traditional scaling operations, which is a modified U-Net \cite{Cicek2015}. We consider both a 2D-CNN and a 3D-CNN implementation for each architecture. Finally, we show an ensemble CNN, which allows combining the longitudinal information leveraged in 3D-CNNs with the proportionally higher value of each segmented voxel in the 2D-CNNs training process, resulting in more accurate results from a theoretical point of view.
The typical assumption behind cross-validation is that the data set is representative of yet to be seen real data, and the test or validation sample should also reflect this. Thus, we would usually balance all folds, so they contain typical samples and also possible outliers. But we want to assess how robust the trained models are and thus we do not randomly mix the folds. Instead, we assign each sample to a fold depending on the number and thickness of its slices. This way, we will always have samples in the test set that are independent of the training data, and we simulate the worst-case scenario for the application in the clinical environment. In order to make the results reproducible, we use open datasets for training and evaluation. We train and validate the CNN-models for kidney tumour segmentation on the dataset of the 2019 Kidney Tumor Segmentation Challenge \cite{Heller2019}. For the liver segmentation, we use the dataset of the CHAOS - Combined (CT-MR) Healthy Abdominal Organ Segmentation Challenge \cite{Selver2019}. %\newline 

It seems like the rise of Deep Learning methods in medical image analysis has split the community into two factions: those who embrace such methods and those who do not trust them. We think that to apply Deep Learning in a clinical setting, the CNN architectures and the entire workflow for data processing and augmentation need to be adapted, requiring considerable knowledge of the diagnostic question and the imaging modality at hand. In this work, we want to show that in order to build clinically applicable CNN-based frameworks, we require different expertise and input from technical and medical domain experts.

\section*{Method}
In the following, we describe the data preprocessing and augmentation in section \ref{preaug}, the network architectures in section \ref{arc} and the training procedure in section \ref{train}. The preprocessing includes volume shape reduction and grey-value windowing. The proposed augmentation addresses the scarcity of data, with the aim of providing additional samples for the training procedure. For the CNN architectures we consider two models: one with dilated convolutions (MS-D) and one with traditional scaling operations (U-Net). We further explain the construction of the stacked CNN model. Subsequently, in section \ref{train} the training procedure for the two considered architectures is described.

\subsection*{Preprocessing and Augmentation}
\label{preaug}
In order to ensure an adequate data quality in the training process for each model, we adapt the data preprocessing and augmentation for CT data. The following description of preprocessing is tailored to the dataset of the KiTS Kidney Tumor Segmentation Challenge \cite{Heller2019} and the dataset of the CHAOS - Combined (CT-MR) Healthy Abdominal Organ Segmentation Challenge \cite{Selver2019}, but can be applied to any other CT dataset with minor changes.

\subsubsection*{Image Preprocessing}
The image normalization is adapted from \cite{Isensee2018} and further extended to make it more general and enable a more realistic normalization for real life applications. 

\begin{figure}[t!]
  \includegraphics[width=\linewidth]{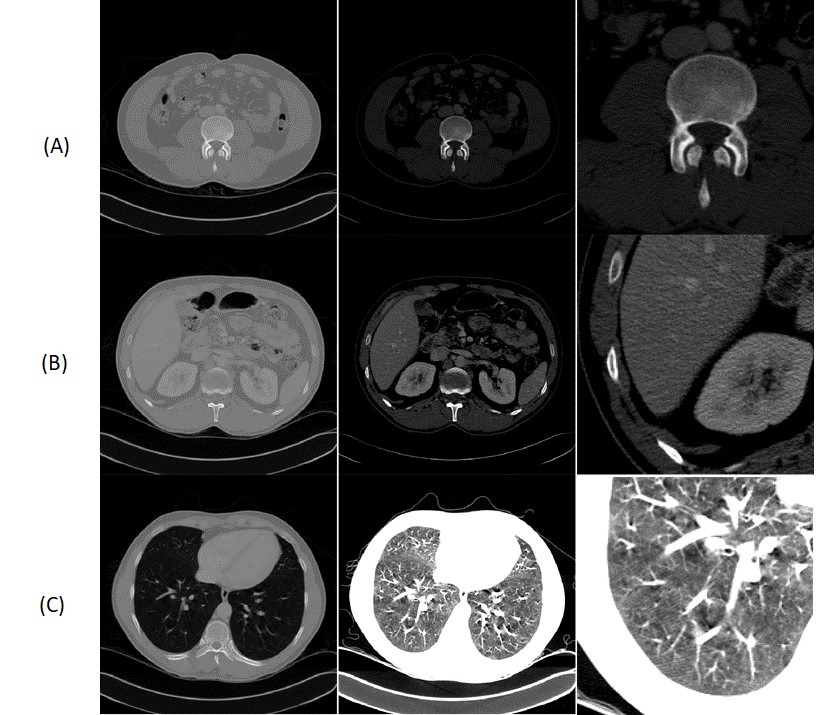}
  \caption{Three examples for the use case oriented windowing ((A) Bone oriented windowing, (B) Organ oriented windowing, (C) Lung oriented windowing). The organ oriented windowing is applied in this work, while the other two examples would be used for the analysis of abnormalities in lung or bony structures in CT.}
\label{fig:windowing}
 \end{figure}

We adapted the image normalisation from \cite{Isensee2018} to better suit real-world applications. To reduce the complexity and optimise the dynamic range, we apply a windowing to each volume by clipping the voxels grey value range to a (0.6, 0.99) percentile range that corresponds to the window a radiologist would use for decision-making. For other segmentation problems, the percentiles must be adjusted to fit the intensity distribution of the relevant body parts (We show examples in \textbf{Figure \ref{fig:windowing}}). We then normalise the windowed data using the z-score using the intensity statistics (mean, standard deviation) from only a random sample of the data set. Using the statistical information from the full dataset would be better but does not reflect the real conditions in a clinical environment. 
 
In order to save costs and time and reduce exposure to radiation in CT  acquisition, the radiologist typically confines a CT acquisition to the region of interest (ROI) (\textbf{Figure \ref{fig:conf_challengs}}). This ROI typically defined liberally not to miss an area that is potentially relevant to the diagnosis. Thus, in a clinical setting, the number of acquired slices in a CT volume considerably varies. This poses a challenge to the application of standard CNN pipelines which often assume a regular data sampling.  To standardise the data, we decided to reduce each volume to 16 slices as we do not need to upsample volumes that contain only a few slices. Instead, our method selects slices at random from each volume, and by repeating the sampling process per volume, we also get a simultaneous data augmentation effect. We exclude background slices during the training phase since these are also not considered in the test phase. We observed that increasing the number of slices did not yield better results, which is consistent with the observation that most CNNs only use a small semantic context for decision making \cite{Hu2017,LaLonde2018}.

In order to save GPU memory, we downsampled each slice from 512 x 512 voxels to 128 x 128 voxels as in our experiments larger slice sizes did not yield better segmentation performance.
 
\subsubsection*{Image Augmentation} 
\begin{figure}[t!]
  \includegraphics[width=\linewidth]{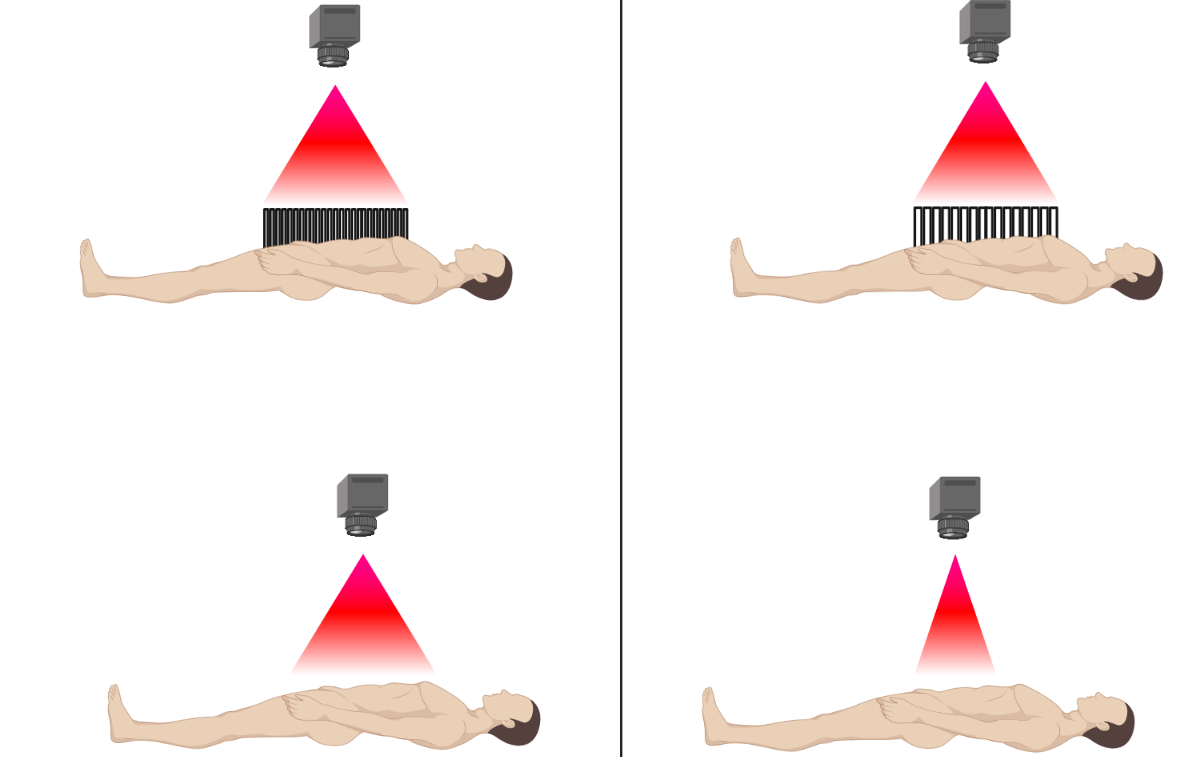}
  \caption{CT scanning configuration, which poses challenges to the application of CNNs. The representation above presents the varying slice thickness, which allows mapping the same region of interest to a different number of slices. The representation below shows the varying size of volumes depending on the chosen region of interest.}
\label{fig:conf_challengs}
 \end{figure}

As additional augmentation steps we used image noising with a normally distributed noise map, slice skipping, slice interpolation and a range shift to address potential variation in the CT acquisition process (\textbf{Figure \ref{fig:conf_challengs}}). We further rotated the images by a random angle (maximum of 16  degrees) to simulate the inevitable variability in patient positioning, that occurs in clinical routine despite fixation. These augmentation steps should more realistically model the expected data variation when applying the deep learning models in clinical practice. 

\subsection*{Architecture}
\label{arc}

To demonstrate the independence of our preprocessing and augmentation framework from the concrete underlying neural network architecture, we compared two conceptually different CNN models. The first architecture we consider here is a modified version of the widely-used U-Net called nnU-Net \cite{Isensee2018}. This architecture extends the original U-Net architecture \cite{Cicek2015} by replacing batch normalization \cite{Ioffe2015} with instance normalization \cite{Ulyanov2016} and ReLUs with LeakyReLU units of slope 1e-2 \cite{Maas2013}. As the second architecture, we chose the mixed-scale dense convolutional neural network (MS-D net) \cite{Pelt2017}. We modified it in the same way as the U-Net to remove the influence of the activation function in our comparison.  We have chosen these two rather extreme variants of CNNs to compare the traditional down- and upscaling flow with the parallel multi-scale approach using dilated convolutions. 

In clinical diagnoses, the radiologist locates the tumour and relevant adjacent structures not only by examining the individual slice but also the adjacent slices. Thus, a 3D CNN might seem like the obvious choice in order to not lose the spatial information from the 3D context. However, previous work has clearly shown that 3D segmentation methods perform worse than 2D approach when the data is anisotropic  \cite{Baumgartner2017,Isensee2017}, which is regularly the case in medical imaging. Another reason why medical image segmentation with 3D CNNs often proves challenging is the variable number of slices per volume. The slice number depends on various external factors like body region under investigation, diagnostic question, different size of the subjects and other trade-offs between data quality, minimal scanning time and radiation exposure. Thus somewhat counterintuitively, 3DD CNNs do not necessarily perform better than 2D versions in many circumstances, and robust models should consider both options.

Here, we combined different models into a single, stacked CNN model to leverage the different strengths of each architecture as ensemble methods showed superior performance in several detection tasks \cite{Dolz17,Kamnitsas18,Teramoto16}. For the kidney-tumour segmentation we stacked a set of 3D MS-D Nets trained to classify voxels into kidney and background (without a distinction between the healthy kidney tissue and the tumour tissue), and a set of 2D nnU-Nets trained to perform classification into healthy tissue, tumour and background. For the liver segmentation, both models perform binary classification of voxels into liver and background.  

\subsection*{Training} 
\label{train}
We trained all networks independently from scratch. The overall training procedure shown in \textbf{Algorithm \ref{training}} was implemented in Python with Tensorflow 1.14 and performed on an IBM Power System Accelerated Compute Server (AC922) with two NVIDIA Tesla V100 GPUs. This setup allowed us to parallelise the experiments, but our proposed approach also works on typical systems with an NVIDIA GTX 1080.
\begin{algorithm}[h]
		\caption{Training procedure}
		\label{training}
		\begin{algorithmic}[1]
			\State Initialize network $f$ with random weights $\theta_{0} $
			\State Initialize validation data $V_{validate}$
			\State Initialize batch size $n$
			\State Assume standard deviation $\sigma$
			\State Select windowing percentile $P$
			\Repeat 
			\Repeat
			\State Select random volume $v$
			\State Windowing($v$, $P_{v}$)
			\State Normalization($v$, $\sigma$)
			\State Augmentation of $v$
			\State Downsampling and slide reduction of $v$
			\State $V_{batch}$ $\leftarrow$  $v$
			\Until {Number of $v$ in $V_{batch}$ = $n$ }
			\State $V_{batch,\hat{y}}$ = $f(V_{batch,x}; \theta_{i})$
			\State $L_i$  = $L_{Tanimoto}$($V_{batch, \hat{y}}$ , $V_{batch, y}$)\textsuperscript{$\alpha$} + $L_{CE}$($V_{batch, \hat{y}}$ , $V_{batch, y}$)\textsuperscript{$\beta$}
			\State $\theta_{i+1}$ = ADAM($L_i$,$\theta_{i}$)
			\State $L_{validation}$  = Validate($f$($V_{validate, x}$;$\theta_{i+1}$,$V_{validate, y}$)
			\Until {Convergence of $L_{validation}$}
		\end{algorithmic}
\end{algorithm}

In each epoch, the volumes of a randomly selected batch are preprocessed and augmented (lines 9-12). We used a batch size of 28 for the 2D networks, while we had to reduce the batch size to 1 (stochastic gradient descent) for the 3D versions of the modified architectures. We use data augmentation in 80 per cent of the training batches for 3D and 90 per cent of training batches in 2D. We applied the intensity range shift to 20 per cent of data in both cases. 

To update the weights $\theta_{i}$ of the neural network function $f$, we used the ADAM optimisation with the parameter configuration proposed in \cite{Kingma2014}. Our loss function $L$ (line 16 in \textbf{Algorithm \ref{training}}) is a combination of the Tanimoto loss $L_{Tanimoto}$ and the categorical crossentropy $L_{CE}$, weighted by $\alpha = 0.6$ and $\beta = 0.4$ respectively. The Tanimoto loss is implemented as shown in equation \ref{eq:score}, where $\hat{y}\in \hat{Y}$ denotes the set of predicted voxel-wise annotations and $y\in Y$ denotes the set of ground truth voxel-wise annotations. The advantage of the Tanimoto coefficient is that it treats each class independently and is thus particularly suitable for problems with a high class imbalance which is typically the case in medical imaging. However, this also leads to a maximum error if a particular class does not occur in the current sample. This effect is attenuated by the smooth factor $smooth$. We empirically chose a small $smooth$ of $1e-5$. A more detailed discussion is given in \cite{Kayalibay2017}. 

\begin{equation}
\label{eq:score}
L_{Tanimoto}(\hat{Y}, Y) = 1 - \frac{\hat{Y}Y + smooth}{\vert\hat{Y}\vert^2+\vert{Y}\vert^2 - \hat{Y}Y + smooth}
\end{equation}

\section*{Evaluation}

\begin{figure}[t!]
  \includegraphics[width=\linewidth]{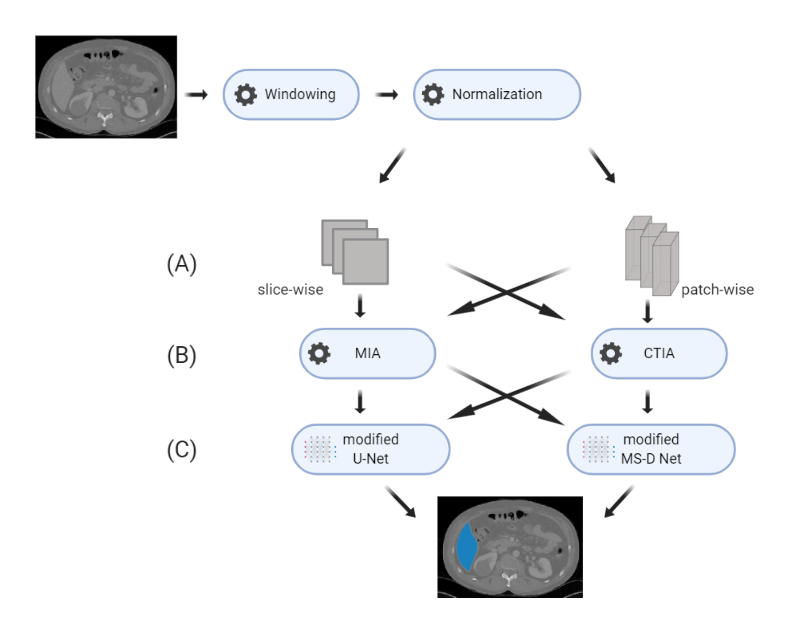}
  \caption{Overview of the workflows considered in the experiments. We switch three parts of the workflow: (A) Input dimensionality, (B) Augmentation toolkit, (C) Convolutional Neural Network. This figure does not include the experiment with the ensemble model.}
\label{fig:experiments}
 \end{figure}

We compared the augmentation of our framework to the multidimensional image augmentation method from \cite{DeepMind18} implemented in TensorFlow  (an illustration of the different experiments is shown in Figure \ref{fig:experiments}). Since the normalisation and the windowing of the CT volume has a strong influence on the cropping and selection of slices, we used the same preprocessing for both augmentation methods. We implemented both CNN architectures in a 2D and 3D version and evaluated each model in a 5-fold cross-validation. To include the influence of edge cases in our validation, we sorted the data according to the number of slices, so the models were always validated on CT volumes that did not occur in the training data set in a similar form. We numerically evaluated the model predictions volume-wise using the Dice score, as shown in equation ~\ref{eq:dice} using the same annotation as in equation ~\ref{eq:score}. We report the resulting scores averaged over volumes and cross-validation folds for the kidney tumour segmentation in table \ref{tab:kidney} and for the liver segmentation in table \ref{tab:liver_results}.

\begin{equation}
\label{eq:dice}
s_{Dice}(\hat{Y}, Y) = \frac{2\hat{Y}Y}{\vert\hat{Y}\vert^2+\vert{Y}\vert^2}
\end{equation}

The results show that the average prediction performance of models trained with CT-specific image augmentation is on par with the performance of models using multidimensional  augmentation. However, the CT-specific preprocessing yields stable results whose standard deviation is an order of magnitude lower than the state-of-the-art multidimensional approach from \cite{DeepMind18}. Our results also confirm the empirical findings that including 3D spatial information in models does not necessarily lead to a better segmentation performance for anisotropic data. \\

\begin{figure}[hpt!]
  \includegraphics[width=\linewidth]{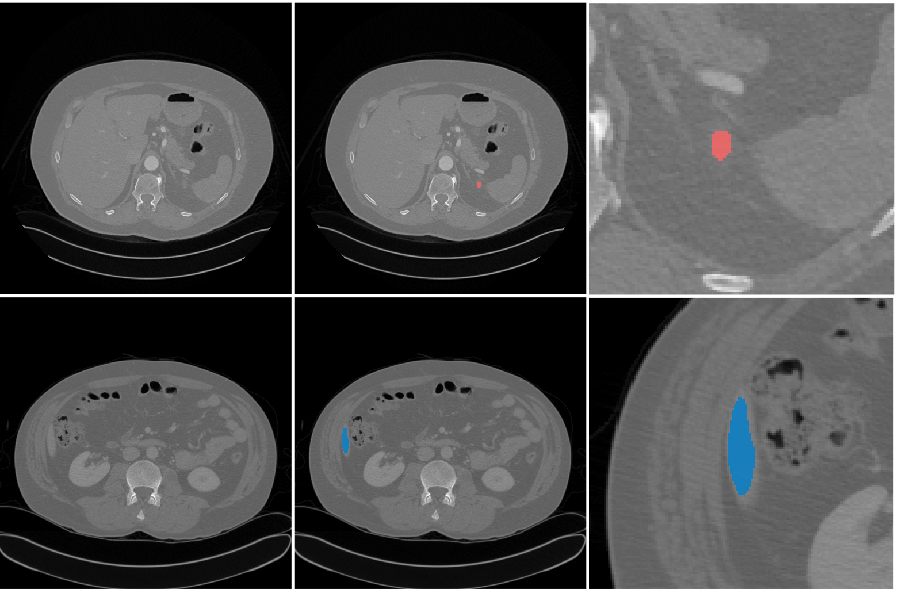}
  \caption{Examples of challenging 2D segmentation cases for liver segmentation (top) and kidney tumor segmentation (bottom). }
  \label{fig:challenge}
\end{figure}

Regarding the different architectures, we found varying results. For the kidney segmentation task, we found that the 3D MS-D Net shows fewer background errors in binary segmentation. These findings indicate that this multi-scale architecture can detect whole objects very well, but the finer distinction between foreground classes (kidney and tumour tissue) works comparatively poorly. For liver segmentation, we found that the MS-D Net generally led to more segmentation errors. However, the MS-D Net errors are typically independent of the segmentation errors of the U-Net approach. In particular, slices with only small regions of interest (shown in \textbf{Figure \ref{fig:challenge}}) pose a challenge. 

Since the errors of the MS-D Net are complementary to the errors of the nnU-Net for both cases, a stacked CNN leads to consistently better results, as it can learn to balance the strengths and weaknesses of the different models. Here, we constructed a stacked CNN consisting of a set of 3D MS-D Nets and a set of 2D nnU-Nets trained with CT-specific image augmentation. For each set, we selected the top-5 models based on their validation score in the previous experiment.  The stacked ensemble of neural network predictor consistently delivered the most accurate and stable predictions by combining the different individual strengths of their members (see table \ref{tab:kidney} and \ref{tab:liver_results}). 

\begin{table}[t]
\caption{Results for the kidney tumor segmentation: Total Dice scores are reported (mean $\pm$ stdv.) for each segmentation class, the different architectures and input dimensionalities (2D and 3D). Each approach is validated with the multidimensional image augmentation (MIA) for Tensorflow and with our CT-specific image augmentation (CTIA).}
\label{tab:kidney}
\centering
\begin{tabular}{|l c   c c c |}
\hline
 & & Kidney & Tumor & Total\\
\hline
nnU-Net + MIA &  2D & $\; 0.962 \pm 0.006 $ & $\; 0.840 \pm 0.013 $ & $\; 0.929 \pm 0.009 $\\
nnU-Net + CTIA &  2D & $\; 0.961 \pm 0.001 $ & $\; 0.844 \pm 0.007 $ & $\; 0.931 \pm 0.002 $\\
 \hline
nnU-Net + MIA &3D & $\; 0.960 \pm 0.012 $ & $\; 0.839 \pm 0.021 $ & $\; 0.929 \pm 0.014 $\\
nnU-Net + CTIA  &3D & $\; 0.960 \pm 0.002 $ & $\; 0.841 \pm 0.008 $ & $\; 0.925 \pm 0.003 $\\
 \hline
MS-D Net + MIA & 2D  & $\; 0.950 \pm 0.011 $ & $\; 0.774 \pm 0.022 $ & $\; 0.913 \pm 0.014 $\\
MS-D Net + CTIA & 2D  & $\; 0.950 \pm 0.001 $ & $\; 0.779 \pm 0.009 $ & $\; 0.914 \pm 0.003 $\\
 \hline
MS-D Net + MIA & 3D &  $\; 0.947 \pm 0.012 $ & $\; 0.764 \pm 0.024 $ & $\; 0.906 \pm 0.018 $\\
MS-D Net + CTIA & 3D &  $\; 0.948 \pm 0.002 $ & $\; 0.765 \pm 0.009 $ & $\; 0.907 \pm 0.003 $\\
\hline
Stacked CNN & & $\; 0.968 \pm 0.001 $ & $\; 0.845 \pm 0.004 $ & $\; 0.947 \pm 0.002 $\\
\hline
\end{tabular}
\end{table}

\begin{table}[h]
\caption{Results for liver segmentation: Total Dice score (mean $\pm$ stdv.) for the different architectures and input dimensionalities (2D and 3D). Each approach is validated with the multidimensional image augmentation (MIA) for Tensorflow and with our CT-specific image augmentation (CTIA).} 

\label{tab:liver_results}
\centering
\begin{tabular}{|l c   c |}
\hline
 & & Total\\
\hline
nnU-Net + MIA &  2D & $\; 0.974 \pm 0.031 $\\
nnU-Net + CTIA &  2D & $\; 0.978 \pm 0.001 $\\
\hline
nnU-Net + MIA &  3D &  $\; 0.941 \pm 0.027 $\\
nnU-Net + CTIA &  3D &  $\; 0.944 \pm 0.014 $\\
 \hline
MS-D Net + MIA & 2D  & $\; 0.961 \pm 0.032 $\\
MS-D Net + CTIA & 2D  & $\; 0.964 \pm 0.002 $\\
\hline
MS-D Net + MIA & 3D &   $\; 0.942 \pm 0.037 $\\
MS-D Net + CTIA & 3D &   $\; 0.942 \pm 0.004 $\\
\hline
Stacked CNN & &  $\; 0.980 \pm 0.001 $\\
\hline
\end{tabular}
\end{table}

\section*{Conclusion}
In this work, we propose a robust machine learning framework for medical image segmentation addressing the specific demands of CT images for clinical applications. Our analysis focused on the often neglected influence of preprocessing and data augmentation on segmentation accuracy and stability. We systematically evaluated this framework for two different state-of-the-art CNN architectures and 2D and 3D input data, respectively. 

In line with previous findings \cite{Baumgartner2017,Isensee2017}, our results show that 3D spatial information does not necessarily lead to better segmentation performance in particular concerning detailed, small-scale image structures. In our experiments, the kind of segmentation errors varied between neural network models, and we showed that a stacked CNN model combining a top-$n$ selection from each model indeed outperformed all other approaches considered in this work. Thus, our findings clearly suggest an ensemble approach as an effective way to achieve more robust and thus, reliable performance in a routine setting. 
Most importantly, our work that our domain-specific data preprocessing scheme yields highly robust segmentation results with an order of magnitude lower variation between samples while maintaining the same average segmentation accuracy as the general-purpose approach independent of the underlying CNN architecture.
 
Existing clinical nephrometry scores have a poor predictive power \cite{Heller2019} and massively reduce the underlying information contained in CT volumes. The improved characterisation of kidney tumours through a more efficient, objective and reliable segmentation, should yield better clinical evaluation, better prediction of clinical outcomes, and ultimately a better treatment of the underlying pathology. In our view, to pave the way to routine clinical applications of machine learning methods for diagnostic decision support, we must focus on improving the robustness and reliability of our segmentation methods. As a first step, our work addresses fundamental methodological challenges in automated segmentation of CT volumes for medical use, to yield reliable organ and tumour segmentation. 

\bibliographystyle{apacite}
\bibliography{sample}

\end{document}